\documentclass[iop,revtex4]{emulateapj}
\usepackage{lineno}

\begin{document}

\renewcommand{\topfraction}{1.0}
\renewcommand{\bottomfraction}{1.0}
\renewcommand{\textfraction}{0.0}

\title{Orbits of subsystems in four hierarchical multiple stars. }

\author{Andrei Tokovinin}
\affil{Cerro Tololo Inter-American Observatory, Casilla 603, La Serena, Chile}
\email{atokovinin@ctio.noao.edu}

\begin{abstract}
Seven  spectroscopic orbits  in nearby  solar-type multiple  stars are
presented.  The primary of the chromospherically active star HIP~9642 is a
4.8-day double-lined  pair; the outer 420-yr visual  orbit is updated,
but  remains poorly  constrained.   HIP 12780  is  a quadruple  system
consisting of the resolved 6.7-yr pair FIN~379 Aa,Ab, for which the combined
orbit,  masses, and  orbital  parallax are  determined  here, and  the
single-lined binary Ba,Bb with a period  of 27.8 days.  HIP 28790 is a
young quintuple system composed of  two close binaries Aa,Ab and Ba,Bb
with periods  of 221 and 13  days, respectively, and  a single distant
component   C.   Its   subsystem   Ba,Bb  is   peculiar,  having   the
spectroscopic  mass  ratio  of  0.89  but a  magnitude  difference  of
$\sim$2.2 mag.  HIP 64478 also contains five stars: the A-component is
a 29-year  visual pair with  a previously known 4-day  twin subsystem,
while the B-component is a contact  binary with a period of 5.8 hours,
seen nearly pole-on.
\end{abstract}


\maketitle

\section{Introduction}
\label{sec:intro}

\begin{deluxetable*}{rr  c l cc rr  c rrr }
\tabletypesize{\scriptsize}     
\tablecaption{Basic parameters of multiple systems
\label{tab:objects} }  
\tablewidth{0pt}                                   
\tablehead{                                                                     
\colhead{HIP} & 
\colhead{HD} & 
\colhead{WDS} & 
\colhead{Spectral} & 
\colhead{$V$} & 
\colhead{$B-V$} & 
\colhead{$\mu^*_\alpha$} & 
\colhead{$\mu_\delta$} & 
\colhead{$\pi_{\rm HIP2}$\tablenotemark{a}}  &
\colhead{$U$} &
\colhead{$V$} &
\colhead{$W$} 
\\
&   &  
\colhead{(J2000)} & 
\colhead{type} & 
\colhead{(mag)} &
\colhead{(mag)} &
\multicolumn{2}{c}{ (mas yr$^{-1}$)} &
\colhead{(mas)} &
\multicolumn{3}{c}{ (km s$^{-1}$)}
}
\startdata
9642  & 12759 & 02039$-$4525 & G5V & 7.31  & 0.69  &+328 & +52  & 20.44 $\pm$ 0.55 & $-$63.4 & $-$59.8 & $-$27.4  \\
12780 & 17134 & 02442$-$2530 & G3V & 6.96  & 0.63  &+166 & +51  & 24.20 $\pm$ 1.16 & $-$27.1 & $-$14.1 & 14.8  \\   
12779 & \ldots& 02442$-$2530 & K0V? & 9.05  & 0.84  &+165 & +52 & 22.86  $\pm$ 1.21 & \ldots & \ldots & \ldots \\   
28790 & 41742 & 06047$-$4505 & F4V  & 5.93 & 0.49 &$-$79 & +255 & 37.18 $\pm$ 0.64 & $-$39.2 & $-$11.0 & $-$14.4  \\  
28764 & 41700 & 06047$-$4505 & F8V  & 6.35 & 0.52 &$-$81  & +246& 37.64 $\pm$ 0.25 & \ldots & \ldots & \ldots  \\  
64478 & 114630& 13129$-$5949 & G0V  & 6.22 & 0.59 &+7  &$-$108  & 23.72  $\pm$ 0.60 & 11.1 & $-$16.1 & $-$20.7  
\enddata
\tablenotetext{a}{Proper motions and parallax from HIP2 \citep{HIP2}.}
\end{deluxetable*}

\begin{figure}
\plotone{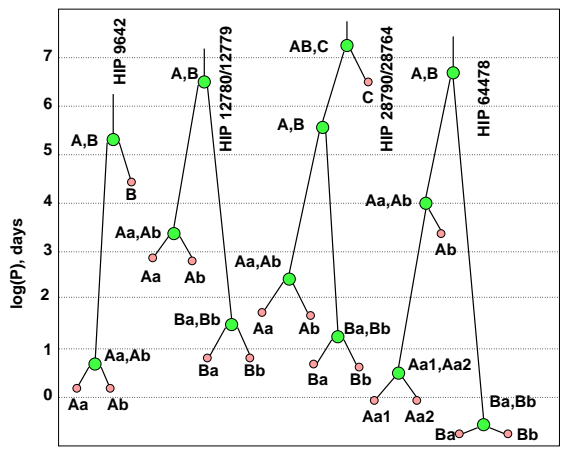}
\caption{
\label{fig:mobile}
Structure of four hierarchical  multiple systems. Larger green circles
depict subsystems (nodes), smaller  pink circles are individual stars.
Vertical  position  of  the   nodes  reflects  orbital  periods  on  a
logarithmic scale.  }
\end{figure}

Radial  velocities (RVs)  of nearby  solar-type multiple  systems were
monitored in 2014 and 2015 to determine the frequency of spectroscopic
subsystems in visually resolved components and to follow their orbital
motion.   The resulting  statistics were  reported  by \citet{survey},
while detailed  analysis of individual systems was  deferred to future
publications.  Four  triple-lined multiples  were featured in  Paper I
\citep{paper1}, and this  is the second paper in  this series. About a
dozen of  remaining stars with  variable RVs observed in  this survey
mostly have  long orbital periods  and still lack sufficient  data for
computing their orbits.

Identifications and  basic data on  the four multiple  systems studied
here are given in Table~1. They have solar-type primary components and
are located  within 50\,pc of the  Sun.  The proper  motions (PMs) and
parallaxes are from \citet{HIP2}, hereafter HIP2.  The spectral types,
$V$  magnitudes, and  $B-V$  colors are  taken  from SIMBAD.   Galactic
velocities $U, V, W$ in the  last three columns are computed using the
HIP2 astrometry  and the center-of-mass  RVs determined here  (the $U$
axis  points away  from the  Galactic center).   The  following common
abbreviations are used throughout this  paper: SB1 and SB2 --- single -
and double-lined  spectroscopic binaries,  AO --- adaptive  optics, PM --- proper motion.
As in Paper  I,  components of multiple systems
are  designated  by single  letters  or  strings,   subsystems  are
denoted   by    their   two   components   joined    with   a   comma.
Figure~\ref{fig:mobile}  depicts  the  hierarchical structure  of  
objects featured  here: one triple,  one quadruple, and  two quintuple
systems.

The observational material and  data analysis are briefly presented in
Section~\ref{sec:obs},     main    results    are     tabulated    in
Section~{\ref{sec:main}.   Sections   4  to  7  are   devoted  to  the
  individual systems.  The paper is summarized in Section~8.

\section{Observations and data analysis}
\label{sec:obs}

The observational  material used here  and the data  reduction methods
are covered in  \citet{survey} and in Paper I.  They are recalled here
briefly.  Most spectra were taken  with the 1.5-m telescope located at
the Cerro  Tololo Interamerican Observatory  in Chile and  operated by
the                    SMARTS                    Consortium.\footnote{
  \url{http://www.astro.yale.edu/smarts/}}  The   observing  time  was
allocated through  NOAO (programs 14B-0009,  15A-0055, 15B-0012).  The
observations were made with  the CHIRON fiber-fed echelle spectrograph \citep{CHIRON} by
the telescope operators in the service mode. HIP~64478 was observed in the
fiber mode with  a spectral resolution of $R=28,000$,  all other stars
were observed in the slicer mode with $R=80,000$.  A few spectra taken
in 2010  at the same  telescope with the  Fiber Echelle (FECH)  with a
resolution of  $R=44,000$ are used, as  well as the  RVs measured with
the echelle spectrometer at the Du Pont 2.5-m telescope \citep{LCO}.

The RVs were derived from  the cross-correlation function (CCF) of the
spectrum with  a binary  mask in the  spectral range from  4500\AA ~to
6500\AA.   The  CCF  is   approximated  by  one  or  several  Gaussian
functions; their centers give  the RVs (after applying the barycentric
correction),  while  the   amplitudes  $a$  and  dispersions  $\sigma$
characterize the depth and width  of the spectral lines.  Paper I gives
approximate  relation between  $\sigma$ and  the  projected rotational
velocity $V \sin i$.

The  orbital  elements  and   their  errors  were  determined  by  the
least-squares fit  to the RVs  with weights inversely  proportional to
the  adopted     errors.   The  IDL   code  {\tt  orbit}\footnote{
  \url{http://www.ctio.noao.edu/\~{}atokovin/orbit/}} was used. It can
fit   spectroscopic,    visual,   or   combined   visual/spectroscopic
orbits. Formal  errors of orbital  elements are determined  from these
fits. For combined  orbits, the errors of orbital  masses and parallax
are computed  by taking  into account correlations  between individual
elements. Masses of  the stars that are not  derived directly from the
orbits, as well as other parameters, are estimated from the absolute
magnitudes using standard relations for main-sequence stars (see Paper
I).

\section{Main results}
\label{sec:main}

\begin{deluxetable}{r l cccc}    
\tabletypesize{\scriptsize}     
\tablecaption{CCF parameters and $V$ magnitudes
\label{tab:CCF}          }
\tablewidth{0pt}                                   
\tablehead{                                                                     
\colhead{HIP} & 
\colhead{Comp.} & 
\colhead{$a$} & 
\colhead{$\sigma$} & 
\colhead{$a \sigma$} & 
\colhead{$V$} \\
&    &   & (km~s$^{-1}$) &  (km~s$^{-1}$) & (mag) 
}
\startdata
9642   & Aa   & 0.262 & 4.49  & 1.174   & 7.70 \\ 
9642   & Ab   & 0.118 & 4.10  & 0.482   & 8.66 \\ 
12780  & Aa   & 0.237 & 3.52 &  0.835 & 7.56 \\
12780  & Ab   & 0.172 & 3.60 &  0.621 & 7.88 \\
12780  & Ba   & 0.429 & 3.30 &  1.570  & 9.05 \\
28790  & Aa   & 0.072 & 14.84 & 1.071 & 6.02 \\
28790  & Ba   & 0.337 &  4.52 & 1.510  & 9.11 \\
28790  & Bb   & 0.051 &  3.93 & 0.197 & 11.32 \\
64478  & Aa1  & 0.067 & 10.19 & 0.681  & 6.99 \\
64478  & Aa2  & 0.068 & 10.20 & 0.689  & 6.99 \\
64478  & Ab   & 0.011 & 5.72  & 0.065  & 10.65 \\
64478  & Ba   & 0.052 & 24.80 &  1.277  & 9.97 \\
64478  & Bb   & 0.043 & 19.53 &  0.851  & 10.42 
\enddata 
\end{deluxetable}

Table~\ref{tab:CCF}   contains   average   parameters  of   the   CCFs
(unresolved CCFs from blended spectra  are not used in the averaging).
The last column lists individual $V$-band magnitudes of the components
computed   from  the   spectroscopic  flux   ratios  (assumed   to  be
proportional to $a \sigma$) and the combined magnitudes.  These estimates 
are accurate to $\sim$0.1 mag and would not be compromised by undetected  small-amplitude 
photometric variability.

Spectroscopic  orbital elements  derived in  this work  are  listed in
Table~\ref{tab:sb},   in common  notation.   Its first two columns contain the {\it Hipparcos} number of the primary component and the subsystem designation. Then follow the orbital period $P$, time of periastron $T$ (for circular orbits it corresponds to the RV maximum), eccentricity $e$, longitude of periastron of the primary component $\omega_{\rm A}$, RV amplitudes $K_1$ and $K_2$  of the primary and secondary components, respectively, and the center-of-mass velocity $\gamma$. The last column gives the weighted rms residuals for both components.  
The   visual    orbits    are   assembled    in
Table~\ref{tab:vb}    ($a$ --- semi-major axis, $\Omega_{\rm A}$ --- position angle of the node, $\omega_{\rm A}$ --- longitude of periastron, $i$ --- inclination). 
The  combined orbit  of HIP
12780  Aa,Ab  is  featured  in both  tables,  duplicating  the overlapping
elements.   In  the  combined  orbit,  the  longitude  of  periastron
$\omega_{\rm A}$ corresponds to  the primary, and the position  angle of the
visual orbit  $\Omega_{\rm A}$ is chosen accordingly to  describe the motion
of the secondary.  


\begin{deluxetable*}{r l cccc ccc c}    
\tabletypesize{\scriptsize}     
\tablecaption{Spectroscopic orbits
\label{tab:sb}          }
\tablewidth{0pt}                                   
\tablehead{                                                                     
\colhead{HIP} & 
\colhead{System} & 
\colhead{$P$} & 
\colhead{$T$} & 
\colhead{$e$} & 
\colhead{$\omega_{\rm A}$ } & 
\colhead{$K_1$} & 
\colhead{$K_2$} & 
\colhead{$\gamma$} & 
\colhead{rms$_{1,2}$} \\
& & \colhead{(d)} &
\colhead{(JD $-$2\,400\,000)} & &
\colhead{(deg)} & 
\colhead{(km~s$^{-1}$)} &
\colhead{(km~s$^{-1}$)} &
\colhead{(km~s$^{-1}$)} &
\colhead{(km~s$^{-1}$)} 
}
\startdata
9642 & Aa,Ab  &      4.78149 & 56970.0145    & 0      &      0    &   27.591 &   31.201      &   49.051 & 0.13 \\
      &       & $\pm$ 0.00002  & $\pm$0.0018 & fixed  & fixed     & $\pm$0.092  & $\pm$0.092 &$\pm$0.04 &  0.12  \\
12780 & Aa,Ab & 2443.1 & 54775.59        & 0.4999 &     191.20 &   7.440 &    8.050 &  $-$0.490 & 0.09 \\ 
     &        &$\pm$2.1 & $\pm$2.5 & $\pm$0.0015 &  $\pm$0.59 & $\pm$0.041 & $\pm$0.041 & $\pm$0.040 & 0.10\\
12780 & Ba,Bb &  27.7679 & 56999.062       & 0.274 &   82.78   &   19.462 &  \ldots &   $-$0.322 & 0.01 \\
     &        &$\pm$0.0033 & $\pm$0.077 & $\pm$0.007 &  $\pm$1.23 & $\pm$0.129 & \ldots & $\pm$0.066 & \ldots\\
28790& Aa,Ab  & 221.385    & 57013.906  & 0.833       &  195.99  &   21.128 &  \ldots &   26.250 & 0.04 \\
     &        & $\pm$0.014 & $\pm$0.218 & $\pm$0.010  &  $\pm$0.46 &   $\pm$1.63 & \ldots & $\pm$0.081 & \ldots \\
28790& Ba,Bb  & 13.2309 & 57004.908     & 0.231       &     123.7   &      21.513 &   24.151 &   28.376 & 0.27 \\
     &        & $\pm$0.0003 & $\pm$0.118 & $\pm$0.017 &   $\pm$3.4  &  $\pm$0.596 & $\pm$0.448 & $\pm$0.218 & 0.39\\
64478 & Aa1,Aa1 & 4.2334536 & 57120.507 & 0  &     0 &      85.172 &   85.364 &   15.364 & 0.22 \\
      &         & $\pm$0.0000018 & $\pm$0.004 & fixed & fixed &    $\pm$   0.40 & $\pm$   0.40 & $\pm$0.237 & 0.18\\
64478 & Ba,Bb  & 0.243524 & 57119.4786 & 0  &  0       &      29.545 &   63.076 &   19.589 & 0.53 \\
      &        &   $\pm$0.000003 & $\pm$0.0004 & fixed & fixed &   $\pm$0.182 & $\pm$0.637 & $\pm$0.160 & 1.83 
\enddata 
\end{deluxetable*}

\begin{deluxetable*}{r l cccc ccc}    
\tabletypesize{\scriptsize}     
\tablecaption{Visual orbits
\label{tab:vb}          }
\tablewidth{0pt}                                   
\tablehead{                                                                     
\colhead{HIP} & 
\colhead{System} & 
\colhead{$P$} & 
\colhead{$T$} & 
\colhead{$e$} & 
\colhead{$a$} & 
\colhead{$\Omega_{\rm A}$ } & 
\colhead{$\omega_{\rm A}$ } & 
\colhead{$i$ }  \\
& & \colhead{(yr)} &
\colhead{(yr)} & &
\colhead{(arcsec)} & 
\colhead{(deg)} & 
\colhead{(deg)} & 
\colhead{(deg)} 
}
\startdata
9642  & A,B   & 415.0   &  2327.57   & 0.2645  &   1.659  &   301.3 &   175.5 &    36.0 \\
12780 & Aa,Ab & 6.68917 &  2008.846  & 0.4999  &   0.1002 &   183.77 &   191.20 &    42.00 \\
      &       & $\pm$0.0051 & $\pm$0.007 & $\pm$0.0015 & $\pm$  0.0006 & $\pm$0.49 & $\pm$0.59 & $\pm$0.72 \\
64478 & Aa,Ab &  29.11  &  1997.27   & 0.173 &   0.3134 &   279.7 &   326.4 &    89.0 \\
      &       & $\pm$0.34 & $\pm$1.55 & $\pm$0.018 & $\pm$0.0034 & $\pm$0.2 & $\pm$17.6 & $\pm$0.4  
\enddata 
\end{deluxetable*}

The observations used in orbit  calculations are listed in two tables,
published in full  electronically.  Table~\ref{tab:rv} gives, for each
date, the RVs of the primary and secondary components $V_1$ and $V_2$,
their errors used  for relative weighting  (unrealistically large errors are assigned to RVs corresponding to blended CCFs), and residuals  to the orbit
O$-$C.   The first  column contains  the {\it  Hipparcos}  number, the
second column  identifies the system.   The dates are given  in
Julian days (minus 2,400,000).  The last column
of  Table~\ref{tab:rv}   specifies  the  data  source.    
The resolved  measurements are listed in  Table~\ref{tab:speckle} in a
similar way  as the  RVs, with the  first two columns  identifying the
system.  Then  follow the  date  in  Besselian  years, position  angle
$\theta$,  separation $\rho$,  position error  $\sigma_\rho$,  residuals to
orbit, and reference.

The following Sections discuss each multiple system individually.

\section{HIP 9642}
\label{sec:9642}

This star is  a visual binary RST~2272.  Its  preliminary visual orbit
by   \citet{Hrt2011d},   revised    here,   has   $P=549.83$\,yr   and
$a$=3\farcs233, leading  to the unrealistically  large mass sum  of 13
${\cal  M}_\odot$.  The  visual secondary  B  is much  fainter than  A
($\Delta Hp  = 4.21$,  $\Delta y  = 4.89$, $\Delta  I \sim  4.4$ mag),
contributing little  light to  the combined spectrum.   Its separation
was last  measured in 2015 at 1\farcs5,  so the light of  B is further
attenuated by the 2\farcs7  entrance aperture of the spectrograph.  It
has no detectable signature in the CCFs.

\begin{figure}
\plotone{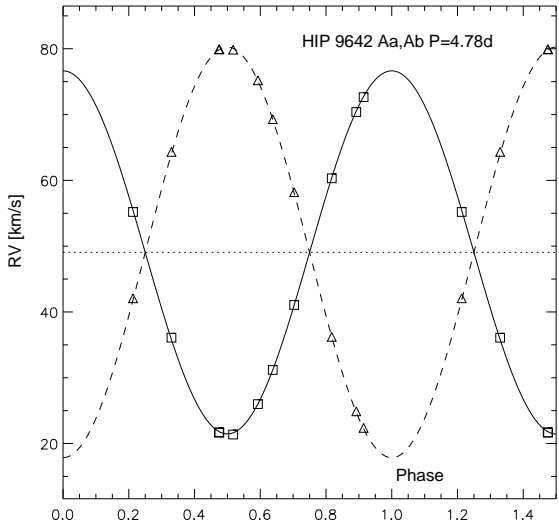}
\caption{Orbit  of  HIP  9642  Aa,Ab  with $P=4.78$\,d.  In  this  and
  following  plots, RVs of  the primary  and secondary  components are
  plotted  as  squares and  triangles,  respectively,  while their  RV
  curves  are traced  by the  full  and dashed  lines. The  horizontal
  dotted line indicates the $\gamma$-velocity.
\label{fig:9642}
}
\end{figure}

The  star  is  a  bright  X-ray  source  and  it  has  an extremely  high
chromospheric   activity   \citep{Makarov2003}.    The   spectroscopic
subsystem  Aa,Ab   was  discovered  by   \citet{N04}.   These  authors
estimated the mass  ratio of 0.86 from nine  observations, but did not
determine the orbit.  Two  observations reported by \citet{LCO} show a
substantial RV variation in one  day, hinting at short period.  The
monitoring  with CHIRON  leads to  a circular  orbit of  Aa,Ab  with a
period  of 4.78  days (Figure~\ref{fig:9642}).   The  eccentricity and
longitude of periastron were fixed at zero in the orbit adjustment.

Using  the  HIP2  parallax, the  magnitudes  of  Aa  and Ab,  and  the
Dartmouth  isochrone  for   solar  metallicity  \citep{Dotter2008},  I
estimated the masses  of Aa and Ab at 1.12  and 0.94 ${\cal M}_\odot$.
The  spectroscopic  orbit  leads  to   $M_1  \sin^3  i$  of  0.05  ${\cal M}_\odot$.
Therefore, the  inclination of the spectroscopic pair  is small, about
22\degr.   The mass  of B  is estimated  at 0.56  ${\cal  M}_\odot$ (a
late-K dwarf), the system mass sum is 2.62 ${\cal M}_\odot$.  However,
the system model based on  the isochrones predicts the $V-K_s$ and $B-V$
colors  of 1.39  and 0.53  mag, respectively,  while the  actual color
indices are 1.78  and 0.69 mag.  So, the stars  are redder than normal
main-sequence dwarfs  (possibly evolved),  and the estimates  of their
masses may be  inaccurate.  The semi-major axis of  Aa,Ab is 1.5\,mas,
so the inner pair can be resolved by long-baseline interferometers.

\begin{figure}
\plotone{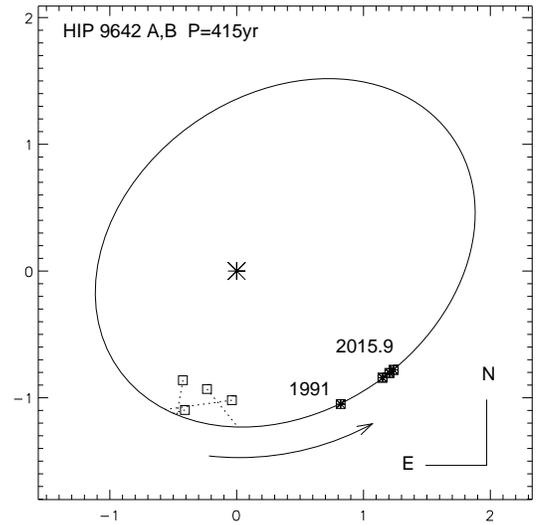}
\caption{New visual  orbit of HIP 9642 A,B  (RST~2272) with $P=415$yr.
  Positions of the secondary component  B relative to the primary A at
  coordinate origin  are plotted as squares (empty  symbols for visual
  data, filled symbols for speckle data). They are connected  by dotted  lines to  the calculated
  positions on the orbital ellipse.  The scale of the axes is in
  arcseconds.   
\label{fig:RST2272}
}
\end{figure}

The   visual  orbit   of  the   outer  system   A,B   lacks  coverage.
Figure~\ref{fig:RST2272}   illustrates  an   alternative   orbit  that
corresponds to the  mass sum of 3.10 ${\cal  M}_\odot$, using the HIP2
parallax.  It was obtained by weighting speckle data more strongly, in
agreement with  their realistic errors,  and by fixing  some elements.
The visual orbit remains provisional  (for this reason no errors are
listed  in Table~\ref{tab:vb}),  but the  new elements  are preferable
because they will give a more accurate ephemeris in the near term. 
The inclination of the outer orbit, 36\degr, if trustworthy, differs
from the estimated inclination of  22\degr ~in the inner orbit.

The  kinematics  (Table~\ref{tab:objects})  and  the  absence  of  the
6708\AA ~lithium  line suggest that  this multiple system is  not very
young.  Its  high chromospheric  activity  likely  results from the tidal
coupling of stellar rotation and orbit.


\section{HIP 12780}
\label{sec:12780}


\begin{deluxetable*}{r l c rrr rrr l}    
\tabletypesize{\scriptsize}     
\tablecaption{Radial velocities and residuals (fragment)
\label{tab:rv}          }
\tablewidth{0pt}                                   
\tablehead{                                                                     
\colhead{HIP} & 
\colhead{System} & 
\colhead{Date} & 
\colhead{$V_1$} & 
\colhead{$\sigma_1$} & 
\colhead{(O$-$C)$_1$ } &
\colhead{$V_2$} & 
\colhead{$\sigma_2$} & 
\colhead{(O$-$C)$_2$ } &
\colhead{Ref.\tablenotemark{a}} \\ 
 & & 
\colhead{(JD $-$2400000)} &
\multicolumn{3}{c}{(km s$^{-1}$)}  &
\multicolumn{3}{c}{(km s$^{-1}$)}  &
}
\startdata
9642 & Aa,Ab &54781.6711 &     36.080 &      0.500 &      0.290 &      64.320 &      0.500 &      0.273 & L \\
9642 & Aa,Ab &54782.5680 &     21.370 &      0.500 &     $-$0.254 &      79.860 &      0.500 &     $-$0.206 & L \\
9642 & Aa,Ab &56896.8724 &     41.060 &      0.200 &      0.029 &      58.201 &      0.200 &      0.080 & C \\
9642 & Aa,Ab &56938.8177 &     21.646 &      0.200 &     $-$0.140 &      79.949 &      0.200 &      0.066 & C \\
12780 & Aa,Ab & 55446.0506 &      0.235 &     10.000 &     $-$2.410  & \ldots   &  \ldots    & \ldots     & F \\
12780 & Aa,Ab & 56908.8456 &     $-$3.938 &      0.050 &     $-$0.161  &    2.846 &      0.050 &     $-$0.179 & C 
\enddata 
\tablenotetext{a}{
C: CHIRON;
F: FECH;
S: \citet{Saar1990};
H: HARPS; 
L: \citet{survey}.
}
\end{deluxetable*}

\begin{deluxetable*}{r l l rrr rr l}    
\tabletypesize{\scriptsize}     
\tablecaption{Position measurements and residuals (fragment)
\label{tab:speckle}          }
\tablewidth{0pt}                                   
\tablehead{                                                                     
\colhead{HIP} & 
\colhead{System} & 
\colhead{Date} & 
\colhead{$\theta$} & 
\colhead{$\rho$} & 
\colhead{$\sigma$} & 
\colhead{(O$-$C)$_\theta$ } & 
\colhead{(O$-$C)$_\rho$ } &
\colhead{Ref.\tablenotemark{a}} \\
 & & 
\colhead{(yr)} &
\colhead{(\degr)} &
\colhead{(\arcsec)} &
\colhead{(\arcsec)} &
\colhead{(\degr)} &
\colhead{(\arcsec)} &
}
\startdata
9642 & A,B &  1932.9100 &    177.9 &    1.020 &    0.500 &     34.9 &   $-$0.196 & Vis  \\
9642 & A,B &  1991.2500 &    218.0 &    1.332 &    0.010 &      1.6 &    0.030 & HIP  \\
9642 & A,B &  2008.7700 &    233.9 &    1.427 &    0.002 &     $-$0.1 &   $-$0.004 & SOAR  \\
12780 & Aa,Ab &  1963.0500 &    142.7 &   0.1140 &   1.0500 &      3.5 &   0.0324 & Fin \\
12780 & Aa,Ab &  1991.7240 &    186.1 &   0.1490 &   0.0020 &      0.2 &   0.0006 & Spe \\
12780 & Aa,Ab &  2015.0280 &    304.4 &   0.0520 &   0.0020 &     $-$1.3 &  $-$0.0014 & SOAR   
\enddata 
\tablenotetext{a}{
Fin: ocular interferometry by W. S. Finsen;
HIP: Hipparcos;
SOAR: speckle interferometry at SOAR;
Spe: speckle interferometry at other telescopes;
Vis: visual micrometer measures.
}
\end{deluxetable*}

\begin{figure}
\plotone{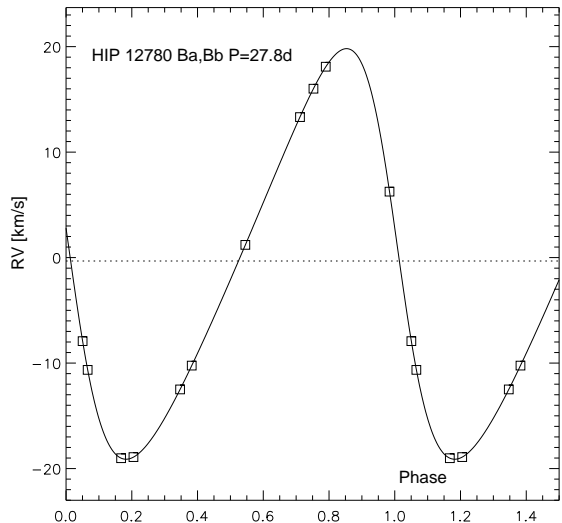}
\caption{Spectroscopic orbit of HIP 12780 Ba,Bb with $P=27.8$\,d.
\label{fig:12780B}
}
\end{figure}

The  primary component  HIP 12780  is a  bright visual  binary FIN~379
Aa,Ab with a  known 6.7-yr orbit   \citep{Hrt2012a}.  The
tertiary component B = HIP~12779  is located at 12\farcs5 from A.  The
two stars have common parallax  and PM.  However, the first
RV  measurement  with  CHIRON  have demonstrated  that  RV(B)  differs
substantially from  RV(A).  Further monitoring  revealed that B  is an
SB1.  Therefore, this multiple system is a 2+2 quadruple.

The  spectroscopic  orbit  of  Ba,Bb  with  $P=27.8$\,d  is  shown  in
Figure~\ref{fig:12780B}. The  very small rms residuals  of 6 m~s$^{-1}$
are  partially explained  by the  fact  that six  orbital elements  are
derived from only 11 RV  measurements.  The estimated mass of Ba, 0.87
${\cal  M}_\odot$, corresponds  to  the minimum  mass  of 0.25  ${\cal
  M}_\odot$ for Bb. It is natural  that Bb is not detected in the CCF.
The component B  was observed with speckle interferometry  at SOAR and
unresolved.

\begin{figure}
\plotone{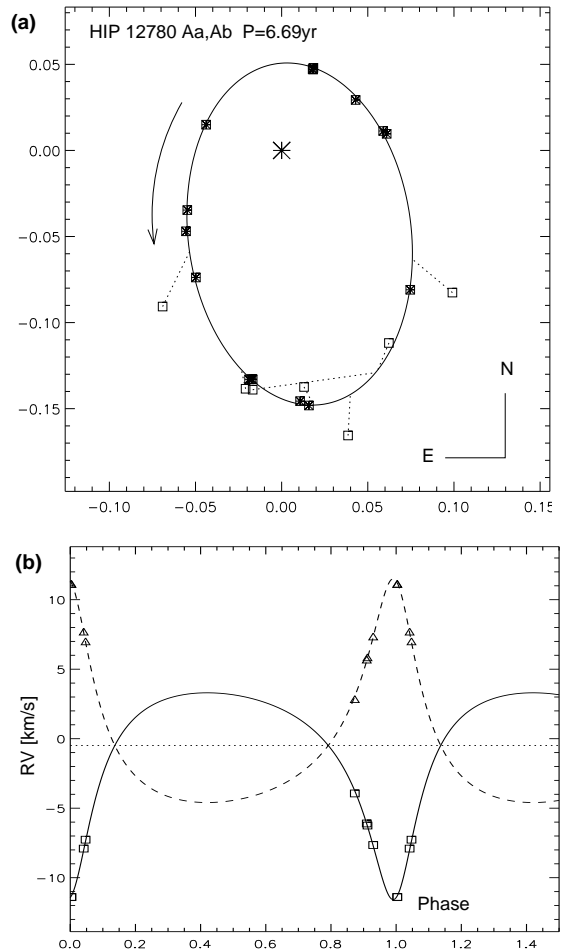}
\caption{Combined orbit of HIP 12780 Aa,Ab = FIN~379.
Top (a): orbit in the sky, bottom (b): the RV curve.
\label{fig:FIN379}
}
\end{figure}

The visual orbit  of the main pair Aa,Ab has  been recently revised by
\citet{Hrt2012a}.  It is in excellent agreement with the RVs of Aa and
Ab  deduced  from  double  CCFs.   Adding  recent  data  from  speckle
interferometry at  SOAR \citep{SAM15},  I computed the  combined orbit
depicted in Figure~\ref{fig:FIN379}.   The weighted rms deviations are
1\fdg2 in angle,  1.2 mas in separation, and 92  and 96 m~s$^{-1}$ for
the RVs of  Aa and Ab, respectively.  The combined  orbit leads to the
masses of 1.05$\pm$0.05 and  0.98$\pm$0.04 ${\cal M}_\odot$ for Aa and
Ab  and the orbital  parallax is  22.26$\pm$0.40\,mas.  Note  that the
parallax of  B = HIP~12779 is  22.9$\pm$1.2 mas, close  to the orbital
parallax of Aa,Ab.   The {\it Hipparcos} parallax of  the main star A,
24.2\,mas,  could   be  slightly biased  by   its  fast  orbital   motion.  The
$\gamma$-velocities of A and B differ by only 0.17 km~s$^{-1}$.

The spectroscopic magnitude difference of Aa,Ab deduced from the areas
of the  CCF dips is  0.32 mag. The  five speckle measures lead  to the
mean $\Delta y = 0.37$ mag, with rms scatter of 0.09 mag. Adopting the
spectroscopic  $\Delta  V$ and  the  orbital  parallax, the  Dartmouth
isochrones \citep{Dotter2008}  lead to the masses that  are 5\% larger
than  the actually  measured ones.   On the  other hand,  the combined
colors of the  component A deduced from the  {\it measured} masses and
the isochrones are in excellent  agreement with the actual colors. The
6708\AA ~lithium line is not detectable  in the spectra of Aa, Ab, and
Ba.

\section{HIP 28790}
\label{sec:28790}

\begin{figure}
\plotone{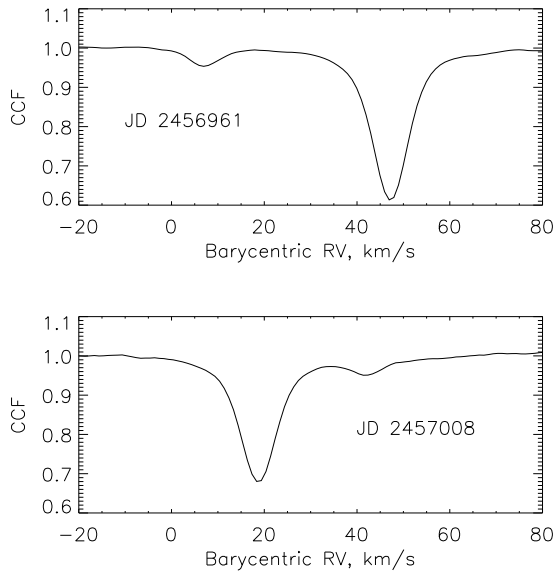}
\caption{CCFs  of HIP~28790B on  two dates showing  double profile
  are plotted on the same  scale.  Note variable amplitude of the main
  dip, 0.37 and 0.30 respectively.
\label{fig:28790B}
}
\end{figure}

\begin{figure}
\plotone{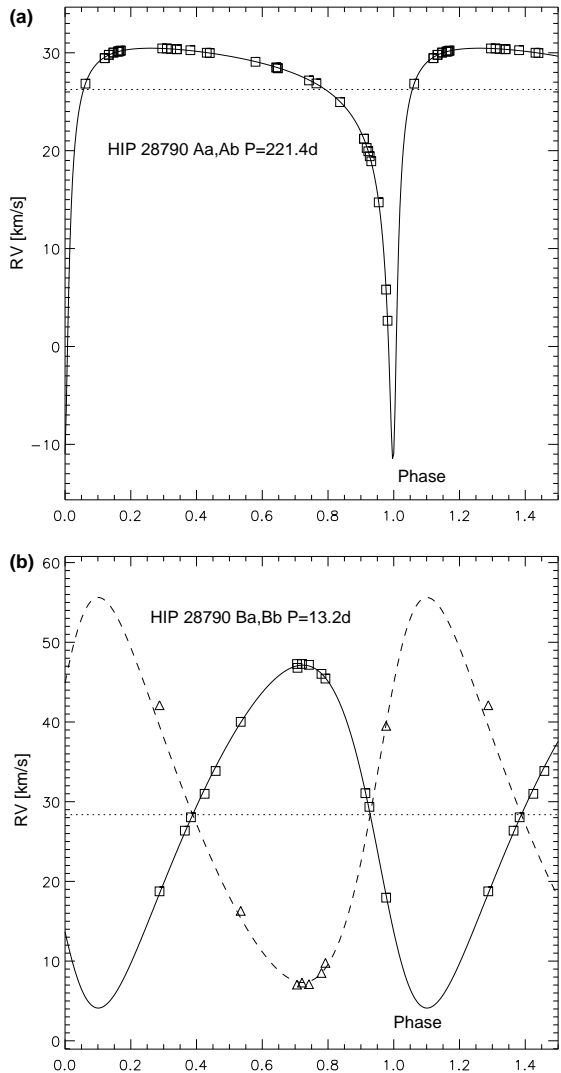}
\caption{Top (a): the spectroscopic orbit  of HIP 28790 Aa,Ab, $P=221.4$d.
  Bottom (b): the RV curve of HIP 28790 Ba,Bb, $P=13.23$\,d.
\label{fig:28790orb}
}
\end{figure}

HIP~28790  is a  young quintuple  stellar system.   The  5\farcs9 pair
HJ~3834 A,B ($V = 6.02; 8.98$) is accompanied by the bright ($V=6.39$)
component C = HIP~28764 at a distance of 196\arcsec.  Both A and B are
spectroscopic  binaries.  The  RV  variation of  A  was discovered  by
\citet{Lagrange2009},  the binarity  of  B was  found by  \citet{LCO}, see  Figure~\ref{fig:28790B}. Spectroscopic orbits of both subsystems are determined here.

Common PM, parallax, and RV  establish the physical nature of the wide
pair  AB,C.   The orbital  periods  of  AB,C  and A,B  estimated  from
projected  separations  are 180\,kyr  and  1  kyr, respectively.   The
binary    A,B    was    observed     with    AO    by    two    groups
\citep{Tok10,Ehrenreich2010}  and  no  additional resolved  components
were  found.    The  component  B  was  also   unresolved  by  speckle
interferometry  at  SOAR.  According  to  the  Washington  Double-Star
Catalog, WDS \citep{WDS}, the pair A,B was at 1\farcs1, 246\degr ~when
it  was discovered in  1837 and  opened up  to 5\farcs9,  215\degr ~in
2010.

The component  A has a fast axial  rotation: the width of  its CCF dip
corresponds to $V \sin i = 26.0$\,km~s$^{-1}$ according to the formula
of   Paper    I.    \citet{Ammler2012}   measured   $V    \sin   i   =
26.7$\,km~s$^{-1}$  and $T_e  =6324$\,K.   \citet{Gray2006} determined
spectral types of F5.5V, K4.5V, and F9V for A, B, and C, respectively,
and  estimated  effective temperatures  of  A  and  C at  6446\,K  and
6241\,K. They found that C is chromospherically active.  This has been
established  earlier   by  \citet{Cutispoto2002},  who   measured  the
rotation of C as $V \sin i = 16.2$\,km~s$^{-1}$.  \citet{Mannings1998}
detected thermal  emission from dust  and identified the main  star as
``Vega-like''.  \citet{Kalas2002} mentioned that the system belongs to
the    $\beta$~Pictoris    group,    but    its    spatial    velocity
(Table~\ref{tab:objects}) does not support  this claim; instead, it is
close  to that of  the Hyades  cluster.  The  spectrum of the component A
contains  the line of  lithium at  6708\AA ~(it  is broadened  by fast
rotation and difficult  to measure), while no such  line is present in
the spectrum of B.

Figure~\ref{fig:28790orb}  (top) shows  the orbital  solution  for the
subsystem  Aa,Ab.  RVs  from CHIRON  are  used together  with the  RVs
measured  by  \citet{Lagrange2009}  with  HARPS.  These  authors  have
kindly  provided  individual  RVs,  not  given in  the  paper,  on  my
request. However, they measured RVs relative to the mean velocity.
 An  offset  of  +29.62\,km~s$^{-1}$  was  found
iteratively to place  those HARPS RVs on the  absolute scale.  The two
data sets together cover well the descending branch of the RV curve in
this  eccentric ($e=0.83$)  orbit,  with  rms residuals  of  only  39\,m~s$^{-1}$
despite fast stellar rotation.  More observations should be planned to
cover the periastron.  If the mass of Aa is  1.2 ${\cal M}_\odot$, the
minimum mass of Ab is 0.47 ${\cal M}_\odot$.

The component B is an SB2. The RV of the main CCF dip is variable, and
sometimes there  is a  weak detail   moving in anti-phase  with the
main dip. Figure~\ref{fig:28790B} shows examples of such CCFs. The SB2
orbit  with   $P=13.2$\,d  is  illustrated  in  the   bottom  plot  of
Figure~\ref{fig:28790orb}.  Approximation of  the CCF by two Gaussians
is not very good, the fit fails in some cases without fixing the width
of the secondary  dip. The rms residuals to the  orbit are larger than
usual.    The  variability  of   the  CCF   amplitude  is   caused  by
contamination  by  the light  of  the  component  A, three  magnitudes
brighter than B and only  at 5\farcs9 distance.  The entrance aperture
of CHIRON has a diameter of  2\farcs7.  Depending on the position of B
on the aperture (while guiding on  the component A) and on the seeing,
a  variable fraction  of  the light  from  A enters  the aperture  and
dilutes the spectrum  of B. The CCFs of  strongly contaminated spectra
contain a wide and weak dip corresponding to the A-component.

The RV amplitudes  of Ba,Bb indicate a mass  ratio $q=0.89$, but the
CCF dips  are very  unequal: the ratio  of their areas  corresponds to
$\Delta V  \approx 2.2$  mag.  The component  Bb must have  a somewhat
later spectral type  than Ba, contributing to the  smaller area of its
CCF dip. Still, the substantial difference  of the CCF areas of Ba and
Bb has no explanation.

The $\gamma$-velocities  of A and  B are 26.25 and  28.38 km~s$^{-1}$,
respectively,  while the  RV  of C  is  27.4~km~s$^{-1}$ and  constant
according to  \citet{N04}. The wide  pair AB,C is  certainly physical,
while the  small RV  difference between  A and B  is explained  by the
orbital motion  of A,B.  The agreement  of RVs makes  it unlikely that
this system contains undiscovered close components, unless they have a
very low mass or a highly inclined orbit. The component C was observed
with the speckle camera at SOAR and unresolved.

\section{HIP 64478}
\label{sec:54478}

\begin{figure}
\plotone{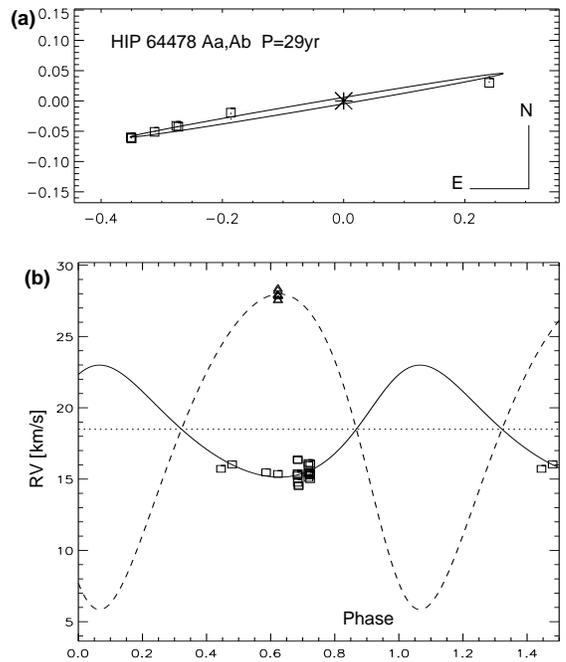}
\caption{Orbit of HIP 64478 Aa,Ab = HDS~1850, $P=29$\,yr.
Top (a): orbit in the plane of the sky, bottom (b): the RV curve. 
\label{fig:HDS1850}
}
\end{figure}

\begin{figure}
\plotone{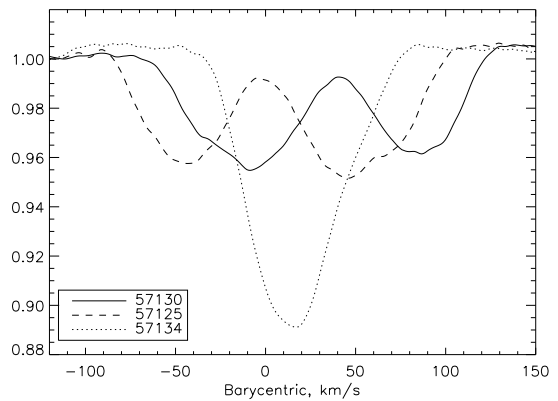}
\caption{CCFs of HIP 64478B on three representative
  dates.
\label{fig:64478corr}
}
\end{figure}

The quintuple system HIP 64478  is also known as HR~4980 or HD~114630.
When  \citet{Saar1990}  determined  the  4.2-day  SB2  orbit  of  this
chromospherically active  G0V binary,  its close visual  companion had
not  yet been  discovered by  {\it Hipparcos},  while the  companion B
located at  25\arcsec ~and 146\degr~ from A  appeared irrelevant. Those
authors wrongly denoted the spectroscopic components as A and B, while
WDS had not yet assigned any designation for B and denoted the visual pair as
COO~152.   The outer system  A,B is  definitely physical,  keeping the
same position since  its discovery in 1892; its  period estimated from
projected  separation   is  about   17\,kyr.  The  PM   of  B   is  
$(+1,-97)$\,mas~yr$^{-1}$, its photometry: $V=9.42$, $K_s=6.68$ mag.

The  0\farcs2  resolved binary  Aa,Ab  is  designated  in the  WDS  as
HDS~1850.  Its preliminary orbit  with a period of 31.6\,yr determined
by \citet{Tok12}  is updated here using  recent speckle interferometry
from SOAR  and the  RVs. So,  the component A  contains in  fact three
stars, all  appearing in the CHIRON  CCFs as distinct  dips.  The weak
dip  corresponding  to  Ab  has  not been  detected  with  CORAVEL  by
\citet{Saar1990}. Considering that B is  now known to be an SB2, there
are five stars in this multiple system.  Several RVs of the components
Aa1 and Aa2  measured with CHIRON are in  excellent agreement with the
SB2 orbit by \citet{Saar1990}.  There is no need to plot this circular
orbit; its elements  derived from the CHIRON data  alone, with a fixed
period, are listed in Table~\ref{tab:sb}.  Combining published and new
data,  the  accurate  period  of  4.2334536 $\pm$  0.0000018  days  is
determined.  With a  mass ratio is 0.998, the  components of this twin
binary are  practically indistinguishable.  \citet{Saar1990} estimated
the orbital  inclination as  85\degr, so there  are no  eclipses.  The
spectroscopic  masses $M \sin^3  i$ of  Aa1 and  Aa2 are  1.085 ${\cal
  M}_\odot$.

The   faint  and   close   companion   Ab  has   an   average  RV   of
27.97\,km~s$^{-1}$  with the  rms  scatter of  0.26 km~s$^{-1}$.   Its
difference  from the $\gamma$-velocity  of Aa,  15.36\,km~s$^{-1}$, is
caused by the motion in the 30-year orbit Aa,Ab.  Using the RVs of Aa1
and  Aa2  measured by  \citet{Saar1990}  with  CORAVEL  and here  with
CHIRON, I computed  the RV of Aa (center of mass  of the inner binary)
as a weighted average, as  explained in Paper I.  Although the CORAVEL
data cover a substantial time span from 1981.1 to 1989.2, only a minor
RV change caused by the motion  in the 30-yr orbit occurred during that
time  (Figure~\ref{fig:HDS1850});  naturally,   no  trend  in  the  RV
residuals of Aa1 and Aa2 has been noted so far.

New observations with CHIRON made 30 years later unfortunately fall on
the  same  orbital  phase, but  contribute  the  RVs  of  Ab.  The  RV  data,
insufficient by themselves, are combined here with relative astrometry
to update  the orbit of Aa,Ab (Figure~\ref{fig:HDS1850}).   It is seen
almost  exactly edge-on.  I  fixed the  $\gamma$-velocity of  Aa,Ab at
18.5  km~s$^{-1}$ to  get the  expected  mass ratio  of $\sim$1/3  and
obtained  the  RV   amplitudes  of  3.93$\pm$0.23  and  11.06$\pm$0.34
km~s$^{-1}$  for  Aa and  Ab,  respectively.   As these  spectroscopic
elements  of  Aa,Ab  are only  a  guess,  they  are not  presented  in
Table~\ref{tab:sb}.

Relative photometry of Aa,Ab at SOAR results in $\Delta y = 3.66$ mag,
with  a 0.06 mag  rms scatter  of the  measures.  As  Aa1 and  Aa2 are
equal, the individual  magnitudes of all three stars  in the aggregate
component A are  estimated from the speckle photometry.   The ratio of
the CCF  areas leads  to $\Delta V_{\rm  Aa,Ab} = 3.30$  mag, slightly
under-estimated because  Ab is cooler than  Aa1 and Aa2  and its lines
are a bit stronger.  The absolute magnitudes imply masses of about 1.2
and 0.7 ${\cal  M}_\odot$ for Aa1 and Ab,  respectively.  However, the
combined  color index  $V-K_s =  1.14$ mag  estimated for  three dwarf
stars of  such masses using the Dartmouth  isochrone differs substantially
from the actual  color $V-K_s = 1.41$ mag.   Apparently the components
of the 4.3-day binary are larger and cooler than normal dwarfs.

\begin{figure}
\plotone{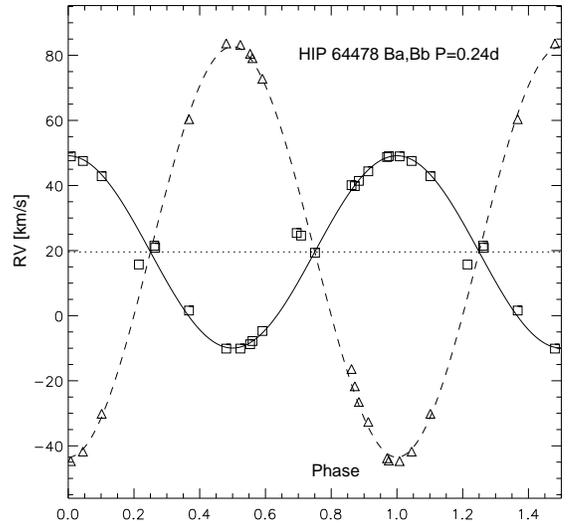}
\caption{RV curve of the subsystem Ba,Bb with period 0.2435 days.
\label{fig:64478orb}
}
\end{figure}

Spectroscopic  observations of  the component  B with  CHIRON  in 2015
revealed  it  as  a double-lined  binary.   Figure~\ref{fig:64478corr}
shows three CCFs of  B plotted on the same scale.  The  dips of Ba and
Bb are obviously widened by  the fast axial rotation. Gaussian fits to
the CCFs are not very accurate. The stronger component Ba has a detail
in its  CCF that moves in anti-phase  with Bb.  The mean  ratio of the
areas of those  Gaussians is 0.69, or $\Delta m =  0.40$ mag, for what
it is worth.

The subsystem Ba,Bb has an unusually short period of 0.2435 days, found
after  several   failed  attempts   to  search  for   longer  periods.
Figure~\ref{fig:64478orb} shows  the RV  curve. Blended CCFs  have not
been used in the  orbit fit, but these RVs are kept  in the plot.  The
rms residuals  to the circular orbit  are rather large,  0.53 and 1.84
km~s$^{-1}$ for Ba and Bb, respectively.

The product $a \sigma$ is a measure of the CCF area.  The sum of areas
of Ba  and Bb  varies between  2.0 and 2.5  km~s$^{-1}$, with  an even
lower value  around 1.6 km~s$^{-1}$  measured on JD  2457121.  Shallow
CCFs correspond to the orbital phases when Ba is approaching and Bb is
receding.  Curiously, the spectrum  of B  has no  H$\alpha$ absorption
line.

The subsystem  Ba,Bb is a contact  binary with a period  of 5.8 hours,
seen almost  from the  pole.  Components of  contact binaries  are not
normal dwarfs and do not follow  standard relations.  The secondary
Bb  is  much  brighter than  follows  from  the  mass ratio  of  0.47.
However, the  complexity of an  interacting binary means that  the RVs
measured here do not reflect the  motion of the center of mass of each
star. 

The combined  colors of  B, $V-K_s =  2.73$ mag  and $B-V =  1.17$ mag,
correspond to a  spectral type around K5V and  imply individual masses
of Ba and Bb of  about 0.7 ${\cal M}_\odot$.  The spectroscopic masses
of  Ba  and Bb  $M\sin^3  i$ are  0.013  and  0.006 ${\cal  M}_\odot$.
Assuming the  mass of Ba to  be 0.7 solar, the  orbital inclination is
$\sim$15\degr. The  equatorial rotation  speed of a  solar-radius star
synchronized with  the orbit is  $\sim$206\,km~s$^{-1}$, the projected
rotational velocity  is then $V \sin i  \approx 52$\,km~s$^{-1}$.  The
stars apparently  rotate synchronously with  the orbit and  fill their
Roche lobes.  The  secondary is smaller and its  projected rotation is
0.8 times less  compared to the primary. The real  mass ratio could be
0.8 or so.  The quintuple system HIP~64478 bears some resemblance to the quintuple doubly-eclipsing
system 1SWASP J093010.78+533859.5 containing low-mass close pairs with periods of 1.31 and 0.25 days 
\citep{Lohr2015}.

I  looked  in the  ASAS  database  \citep{ASAS}  to test  whether  the
component B  could be  variable. Unfortunately, there  are only  a few
measurements of B there. One night of photometric monitoring is needed
to determine the amplitude or to set its upper limit.


\section{Discussion and summary}
\label{sec:sum}

This  work  started from  monitoring  secondary  components of  nearby
solar-type stars in order to determine the frequency and parameters of
subsystems.   The   new  orbits  given   here  strengthen  statistical
knowledge  of low-mass binaries  in the  solar vicinity,  reducing the
number  of subsystems  with unknown  periods.  The  primary components
were measured as well for  a consistency check and yielded interesting
results, including the first  orbit of HIP~28790 Aa,Ab, measurement of
masses of  HIP~12780 Aa,Ab, and the  orbit of Aa,Ab  in the three-tier
hierarchy  HIP~64478.  In  contrast, the  subsystem Aa,Ab  in HIP~9642
belongs to  the primary  component, while the  faint and  close visual
secondary is not detected spectroscopically, making it a rather common
and uninteresting triple. 

Two secondary subsystems featured here are noteworthy. HIP~28790 Ba,Bb
is  an SB2  with a  mass ratio  of 0.89,  yet very  unequal components
(Figure~\ref{fig:28790B}).    There   is   no  explanation   of   this
paradox. This multiple system may be relatively young, as evidenced by
its kinematics, the rotation of Aa, and the presence of lithium in its
photosphere. 

The binary HIP  64478 Ba,Bb is even more exotic,  being a contact pair
of 5.8\,h period observed at low orbital inclination of $\sim$15\degr.
This  is  probably  a  unique  case, as  other  contact  binaries  are
discovered  photometrically by eclipses  or by  ellipsoidal variation.
Contact  binaries are  relatively rare,  about  one per  500 stars  of
spectral types F, G, K \citep{Rucinski2002,Rucinski2006}.  Compared to
the majority of  known contact binaries, HIP 64478  Ba,Bb has a rather
short period  and a faint absolute  magnitude of $M_V =  6.3$ mag.  As
its  distance  is well  known,  it  can  improve the  currently  known
relation   between   period  and   absolute   magnitude  for   contact
systems.   The  sample   of  4847   solar-type  stars   within  67\,pc
\citep{FG67b} contains 14 eclipsing  systems with periods shorter than
one day.  The  period of  HIP 64478  Ba,Bb is the  shortest in  the whole
sample.  This  bright, nearby,  and unusually oriented  system, if
studied  in  greater detail,  may  provide  interesting insights  into
physics of such merging pairs.

Looking at the mobile diagrams  in Figure~\ref{fig:mobile}, we notice that
when both  visual components contain  close subsystems, the  period of
the more  massive primary is longer  than the period  of the secondary
subsystem. Such  a trend is expected from the general correlation between
mass and angular momentum. It should be studied on a larger sample, as
the  evidence from only  three cases  is circumstantial.   

Yet another feature of this diagram  is a large ($\sim 10^3$) ratio of
periods  at the  lowest hierarchical  levels (i.e.   a  large vertical
distance  between the lowest  and the  next nodes).   This could  be a
combined result of the  formation process, where inner subsystems form
first and  shrink rapidly before  acquiring outer companions,  and the
posterior dynamical evolution involving tidal friction and Kozai-Lidov
cycles \citep{KCTF}.  This latter  process works when inner pairs have
separations of a  few solar radii at periastron and  the orbits have a
large mutual inclination. It leaves a population of triples with inner
periods  of a  few  days and  relative  inclinations clustered  around
39\degr  ~and  141\degr.   The  inner binaries  continue  to  interact
tidally  and their orbits  are rapidly  circularized.  HIP  9642 Aa,Ab
could result from  such evolution, but not all  inner subsystems 
  match  this scenario,  being  too  wide  and/or having  eccentric
orbits. So, tidal evolution cannot be a unique way to form close inner
subsystems and  some of them should be  primordial.  Considering large
period ratios, formation of close subsystems by dynamical interactions
in unstable  small groups  of nascent stars  is unlikely,  unless they
evolved subsequently to much shorter periods.


\acknowledgements

I  thank   the  operators  of   the  1.5-m  telescope   for  executing
observations  of  this  program  and  the  SMARTS  team  at  Yale  for
scheduling  and pipeline  processing.  Archival  measures  of resolved
binaries were  retrieved from the  Washington database by  B.~Mason. I
thank A.-M.~Lagrange and  S.~Borgniet for communicating individual RVs
of  HIP  28790A, S.~Rucinski  for  commenting  upon  the close  binary
HIP~64478  Ba,Bb and  reading the  paper  draft. 
 Referee's comments helped to improve the paper.  This  work used  the
SIMBAD   service   operated  by   Centre   des  Donn\'ees   Stellaires
(Strasbourg, France),  bibliographic references from  the Astrophysics
Data System maintained by SAO/NASA, the Washington Double Star Catalog
maintained at USNO, and products of the 2MASS survey.

{\it Facilities:}  \facility{CTIO:1.5m}, \facility{SOAR} 




\clearpage

\LongTables
\setcounter{table}{4}


\begin{deluxetable*}{r l c rrr rrr l}    
\tabletypesize{\scriptsize}     
\tablecaption{Radial velocities and residuals
\label{tab:rv}          }
\tablewidth{0pt}                                   
\tablehead{                                                                     
\colhead{HIP} & 
\colhead{System} & 
\colhead{Date} & 
\colhead{$V_1$} & 
\colhead{$\sigma_1$} & 
\colhead{(O$-$C)$_1$ } &
\colhead{$V_2$} & 
\colhead{$\sigma_2$} & 
\colhead{(O$-$C)$_2$ } &
\colhead{Ref.\tablenotemark{a}} \\ 
 & & 
\colhead{(JD $-$2,400,000)} &
\multicolumn{3}{c}{(km s$^{-1}$)}  &
\multicolumn{3}{c}{(km s$^{-1}$)}  &
}
\startdata
9642 & Aa,Ab &54781.6711 &     36.080 &      0.500 &      0.290 &      64.320 &      0.500 &      0.273 & L \\
9642 & Aa,Ab &54782.5680 &     21.370 &      0.500 &     -0.254 &      79.860 &      0.500 &     -0.206 & L \\
9642 & Aa,Ab &56896.8724 &     41.060 &      0.200 &      0.029 &      58.201 &      0.200 &      0.080 & C \\
9642 & Aa,Ab &56938.8177 &     21.646 &      0.200 &     -0.140 &      79.949 &      0.200 &      0.066 & C \\
9642 & Aa,Ab &56967.5011 &     21.707 &      0.200 &     -0.111 &      79.893 &      0.200 &      0.046 & C \\
9642 & Aa,Ab &56969.6052 &     72.647 &      0.200 &     -0.100 &      22.373 &      0.200 &      0.117 & C \\
9642 & Aa,Ab &56977.6304 &     25.987 &      0.200 &     -0.031 &      75.204 &      0.200 &      0.106 & C \\
9642 & Aa,Ab &56978.7048 &     60.310 &      0.200 &     -0.093 &      36.206 &      0.200 &     -0.008 & C \\
9642 & Aa,Ab &56980.5951 &     55.197 &      0.200 &     -0.240 &      42.070 &      0.200 &      0.241 & C \\
9642 & Aa,Ab &56982.6278 &     31.182 &      0.200 &     -0.009 &      69.300 &      0.200 &      0.052 & C \\
9642 & Aa,Ab &56988.6238 &     70.384 &      0.200 &     -0.140 &      24.887 &      0.200 &      0.118 & C \\
12780 & Aa,Ab & 55446.0506 &      0.235 &     10.000 &     -2.410  & \ldots   &  \ldots    & \ldots     & F \\
12780 & Aa,Ab & 56908.8456 &     -3.938 &      0.050 &     -0.161  &    2.846 &      0.050 &     -0.179 & C \\
12780 & Aa,Ab & 56998.6952 &     -6.098 &      0.050 &      0.066  &    5.690 &      0.050 &      0.078 & C \\
12780 & Aa,Ab & 57003.8085 &     -6.249 &      0.050 &      0.073  &    5.854 &      0.050 &      0.071 & C \\
12780 & Aa,Ab & 57046.5419 &     -7.649 &      0.050 &      0.072  &    7.354 &      0.050 &      0.055 & C \\
12780 & Aa,Ab & 57226.9715 &    -11.383 &      0.050 &     -0.066  &   11.150 &      0.050 &     -0.046 & C \\
12780 & Aa,Ab & 57226.9715 &    -11.400 &      0.050 &     -0.083  &   11.119 &      0.050 &     -0.077 & C \\
12780 & Aa,Ab & 57319.7430 &     -7.905 &      0.050 &      0.110  &    7.705 &      0.050 &      0.088 & C \\
12780 & Ba,Bb & 55445.8833 &    -10.646 &      2.000 &     -0.016 &   \ldots   &  \ldots    & \ldots     & F \\
12780 & Ba,Bb & 56908.8883 &     16.010 &      0.200 &     -0.003 &   \ldots   &  \ldots    & \ldots     & C \\
12780 & Ba,Bb & 56937.6987 &     18.094 &      0.200 &      0.009 &   \ldots   &  \ldots    & \ldots     & C \\
12780 & Ba,Bb & 56998.6192 &      6.252 &      0.200 &     -0.002 &   \ldots   &  \ldots    & \ldots     & C \\
12780 & Ba,Bb & 57003.7136 &    -19.018 &      0.200 &     -0.003 &   \ldots   &  \ldots    & \ldots     & C \\
12780 & Ba,Bb & 57008.6948 &    -12.497 &      0.200 &     -0.001 &   \ldots   &  \ldots    & \ldots     & C \\
12780 & Ba,Bb & 57009.6803 &    -10.238 &      0.200 &     -0.003 &   \ldots   &  \ldots    & \ldots     & C \\
12780 & Ba,Bb & 57046.5915 &     13.321 &      0.200 &     -0.010 &   \ldots   &  \ldots    & \ldots     & C \\
12780 & Ba,Bb & 57226.9011 &    -18.911 &      0.200 &      0.003 &   \ldots   &  \ldots    & \ldots     & C \\
12780 & Ba,Bb & 57319.6545 &      1.207 &      0.200 &      0.008 &   \ldots   &  \ldots    & \ldots     & C \\
12780 & Ba,Bb & 57333.6777 &     -7.920 &      0.200 &      0.002 &   \ldots   &  \ldots    & \ldots     & C \\
28790 & Aa,Ab & 53989.8028 &     30.368 &      0.100 &      0.016 &    \ldots   &  \ldots    & \ldots     & H \\
28790 & Aa,Ab & 53989.8067 &     30.357 &      0.100 &      0.005 &    \ldots   &  \ldots    & \ldots     & H \\
28790 & Aa,Ab & 54056.8309 &     28.500 &      0.100 &     -0.020 &    \ldots   &  \ldots    & \ldots     & H \\
28790 & Aa,Ab & 54056.8348 &     28.494 &      0.100 &     -0.026 &    \ldots   &  \ldots    & \ldots     & H \\
28790 & Aa,Ab & 54057.7883 &     28.421 &      0.100 &     -0.054 &    \ldots   &  \ldots    & \ldots     & H \\
28790 & Aa,Ab & 54057.7924 &     28.420 &      0.100 &     -0.055 &    \ldots   &  \ldots    & \ldots     & H \\
28790 & Aa,Ab & 54386.7594 &     29.790 &      0.100 &      0.019 &    \ldots   &  \ldots    & \ldots     & H \\
28790 & Aa,Ab & 54386.7650 &     29.798 &      0.100 &      0.026 &    \ldots   &  \ldots    & \ldots     & H \\
28790 & Aa,Ab & 54389.8641 &     30.022 &      0.100 &      0.054 &    \ldots   &  \ldots    & \ldots     & H \\
28790 & Aa,Ab & 54389.8749 &     30.022 &      0.100 &      0.053 &    \ldots   &  \ldots    & \ldots     & H \\
28790 & Aa,Ab & 54392.8366 &     30.125 &      0.100 &      0.014 &    \ldots   &  \ldots    & \ldots     & H \\
28790 & Aa,Ab & 54392.8416 &     30.125 &      0.100 &      0.013 &    \ldots   &  \ldots    & \ldots     & H \\
28790 & Aa,Ab & 54393.7776 &     30.151 &      0.100 &      0.002 &    \ldots   &  \ldots    & \ldots     & H \\
28790 & Aa,Ab & 54393.7829 &     30.155 &      0.100 &      0.006 &    \ldots   &  \ldots    & \ldots     & H \\
28790 & Aa,Ab & 54394.7703 &     30.213 &      0.100 &      0.027 &    \ldots   &  \ldots    & \ldots     & H \\
28790 & Aa,Ab & 54394.7754 &     30.220 &      0.100 &      0.034 &    \ldots   &  \ldots    & \ldots     & H \\
28790 & Aa,Ab & 54422.7526 &     30.445 &      0.100 &      0.002 &    \ldots   &  \ldots    & \ldots     & H \\
28790 & Aa,Ab & 54425.8466 &     30.432 &      0.100 &      0.012 &    \ldots   &  \ldots    & \ldots     & H \\
28790 & Aa,Ab & 54428.7658 &     30.384 &      0.100 &     -0.010 &    \ldots   &  \ldots    & \ldots     & H \\
28790 & Aa,Ab & 54452.7270 &     30.008 &      0.100 &     -0.009 &    \ldots   &  \ldots    & \ldots     & H \\
28790 & Aa,Ab & 54485.5476 &     29.080 &      0.100 &     -0.017 &    \ldots   &  \ldots    & \ldots     & H \\
28790 & Aa,Ab & 54521.5978 &     27.177 &      0.100 &     -0.044 &    \ldots   &  \ldots    & \ldots     & H \\
28790 & Aa,Ab & 54542.5150 &     24.958 &      0.100 &     -0.046 &    \ldots   &  \ldots    & \ldots     & H \\
28790 & Aa,Ab & 54558.5899 &     21.228 &      0.100 &      0.039 &    \ldots   &  \ldots    & \ldots     & H \\
28790 & Aa,Ab & 54558.5937 &     21.212 &      0.100 &      0.025 &    \ldots   &  \ldots    & \ldots     & H \\
28790 & Aa,Ab & 54781.8038 &     20.280 &      0.500 &     -0.156 &    \ldots   &  \ldots    & \ldots     & L \\
28790 & Aa,Ab & 56961.8068 &     26.872 &      0.100 &      0.053 &    \ldots   &  \ldots    & \ldots     & C \\
28790 & Aa,Ab & 56996.7491 &     19.937 &      0.100 &      0.011 &    \ldots   &  \ldots    & \ldots     & C \\
28790 & Aa,Ab & 56997.7431 &     19.427 &      0.100 &      0.012 &    \ldots   &  \ldots    & \ldots     & C \\
28790 & Aa,Ab & 56998.7520 &     18.917 &      0.100 &      0.074 &    \ldots   &  \ldots    & \ldots     & C \\
28790 & Aa,Ab & 57003.7703 &     14.719 &      0.100 &     -0.030 &    \ldots   &  \ldots    & \ldots     & C \\
28790 & Aa,Ab & 57008.7167 &      5.810 &      0.100 &     -0.058 &    \ldots   &  \ldots    & \ldots     & C \\
28790 & Aa,Ab & 57009.7352 &      2.627 &      0.100 &      0.032 &    \ldots   &  \ldots    & \ldots     & C \\
28790 & Aa,Ab & 57027.6661 &     26.845 &      0.100 &     -0.013 &    \ldots   &  \ldots    & \ldots     & C \\
28790 & Aa,Ab & 57040.7020 &     29.458 &      0.100 &     -0.092 &    \ldots   &  \ldots    & \ldots     & C \\
28790 & Aa,Ab & 57261.9109 &     29.440 &      0.100 &     -0.094 &    \ldots   &  \ldots    & \ldots     & C \\
28790 & Aa,Ab & 57319.6913 &     30.265 &      0.100 &      0.043 &    \ldots   &  \ldots    & \ldots     & C \\
28790 & Aa,Ab & 57332.7685 &     29.977 &      0.100 &      0.003 &    \ldots   &  \ldots    & \ldots     & C \\
28790 & Ba,Bb & 54781.8075 &     17.960 &      0.500 &     -0.291 &      39.520 &      0.500 &     -0.223 & L \\
28790 & Ba,Bb & 56934.8785 &     46.794 &      0.500 &     -0.266 &     \ldots  &  \ldots    & \ldots     & C \\
28790 & Ba,Bb & 56937.7719 &     29.357 &      0.500 &      0.475 &     \ldots  &  \ldots    & \ldots     & C \\
28790 & Ba,Bb & 56961.8014 &     47.172 &      0.500 &      0.142 &       7.109 &      0.500 &     -0.326 & C \\
28790 & Ba,Bb & 56996.7440 &     28.044 &      0.500 &      0.163 &     \ldots  &  \ldots    & \ldots     & C \\
28790 & Ba,Bb & 56997.7384 &     33.856 &      0.500 &     -0.516 &     \ldots  &  \ldots    & \ldots     & C \\
28790 & Ba,Bb & 56998.7472 &     40.022 &      0.500 &      0.096 &      16.304 &      5.000 &      0.894 & C \\
28790 & Ba,Bb & 57003.7650 &     31.066 &      0.500 &     -0.177 &     \ldots  &  \ldots    & \ldots     & C \\
28790 & Ba,Bb & 57008.7114 &     18.755 &      0.500 &      0.270 &      42.106 &      5.000 &      2.625 & C \\
28790 & Ba,Bb & 57009.7300 &     26.368 &      0.500 &      0.222 &     \ldots  &  \ldots    & \ldots     & C \\
28790 & Ba,Bb & 57027.6764 &     47.290 &      0.500 &      0.158 &       7.304 &      0.500 &     -0.017 & C \\
28790 & Ba,Bb & 57040.7063 &     47.273 &      0.500 &      0.227 &       7.051 &      0.500 &     -0.366 & C \\
28790 & Ba,Bb & 57261.9131 &     30.987 &      1.000 &     -0.605 &     \ldots  &  \ldots    & \ldots     & C \\
28790 & Ba,Bb & 57319.7004 &     45.449 &      0.500 &     -0.075 &       9.773 &      0.500 &      0.647 & C \\
28790 & Ba,Bb & 57332.7705 &     46.043 &      0.500 &     -0.032 &       8.514 &      1.000 &      0.006 & C \\
64478 & Aa1,Aa2 & 57141.6159 &    100.176 &      1.000 &      0.037 &     -69.700 &      1.000 &     -0.177 & C \\
64478 & Aa1,Aa2 & 57162.5490 &     92.760 &      1.000 &     -0.129 &     -61.948 &      1.000 &      0.301 & C \\
64478 & Aa1,Aa2 & 57162.6056 &     95.449 &      1.000 &     -0.094 &     -64.623 &      1.000 &      0.289 & C \\
64478 & Aa1,Aa2 & 57166.5739 &     78.058 &      1.000 &     -0.527 &     -47.751 &      1.000 &      0.146 & C \\
64478 & Aa1,Aa2 & 57166.5760 &     78.336 &      1.000 &     -0.426 &     -47.946 &      1.000 &      0.128 & C \\
64478 & Aa1,Aa2 & 57169.5730 &    -56.343 &      1.000 &     -0.610 &      87.491 &      1.000 &      0.612 & C \\
64478 & Aa1,Aa2 & 57170.5171 &     47.611 &      1.000 &     -1.484 &     -17.454 &      1.000 &      0.852 & C \\
64478 & Aa1,Aa2 & 57175.6150 &    100.178 &      1.000 &      0.402 &     -69.357 &      1.000 &     -0.198 & C \\
64478 & Aa1,Aa2 & 57177.4994 &    -67.318 &      1.000 &      0.013 &      98.407 &      1.000 &     -0.109 & C \\
64478 & Aa1,Aa2 & 57177.6185 &    -69.709 &      1.000 &     -0.331 &     100.414 &      1.000 &     -0.156 & C \\
64478 & Aa,Ab & 44646.5692 &     15.717 &      0.500 &     -0.178 &\ldots     & \ldots & \ldots & S \\
64478 & Aa,Ab & 45016.9248 &     16.026 &      0.500 &      0.301 &\ldots     & \ldots & \ldots & S \\
64478 & Aa,Ab & 46142.9665 &     15.470 &      0.500 &      0.089 &\ldots     & \ldots & \ldots & S \\
64478 & Aa,Ab & 47189.7507 &     15.248 &      0.500 &     -0.122 &\ldots     & \ldots & \ldots & S \\
64478 & Aa,Ab & 47190.8464 &     15.416 &      0.500 &      0.046 &\ldots     & \ldots & \ldots & S \\
64478 & Aa,Ab & 47191.9421 &     16.380 &      0.500 &      1.010 &\ldots     & \ldots & \ldots & S \\
64478 & Aa,Ab & 47193.0378 &     15.322 &      0.500 &     -0.049 &\ldots     & \ldots & \ldots & S \\
64478 & Aa,Ab & 47193.7683 &     14.742 &      0.500 &     -0.629 &\ldots     & \ldots & \ldots & S \\
64478 & Aa,Ab & 47194.8641 &     16.326 &      0.500 &      0.955 &\ldots     & \ldots & \ldots & S \\
64478 & Aa,Ab & 47220.7962 &     14.576 &      0.500 &     -0.800 &\ldots     & \ldots & \ldots & S \\
64478 & Aa,Ab & 47221.8920 &     14.500 &      0.500 &     -0.876 &\ldots     & \ldots & \ldots & S \\
64478 & Aa,Ab & 47545.8618 &     15.987 &      0.500 &      0.527 &\ldots     & \ldots & \ldots & S \\
64478 & Aa,Ab & 47546.9575 &     15.082 &      0.500 &     -0.378 &\ldots     & \ldots & \ldots & S \\
64478 & Aa,Ab & 47547.6880 &     15.397 &      0.500 &     -0.063 &\ldots     & \ldots & \ldots & S \\
64478 & Aa,Ab & 47548.7837 &     16.128 &      0.500 &      0.667 &\ldots     & \ldots & \ldots & S \\
64478 & Aa,Ab & 47549.8795 &     15.479 &      0.500 &      0.018 &\ldots     & \ldots & \ldots & S \\
64478 & Aa,Ab & 47602.8396 &     15.241 &      0.500 &     -0.238 &\ldots     & \ldots & \ldots & S \\
64478 & Aa,Ab & 47603.9353 &     15.383 &      0.500 &     -0.097 &\ldots     & \ldots & \ldots & S \\
64478 & Aa,Ab & 47604.6658 &     14.991 &      0.500 &     -0.489 &\ldots     & \ldots & \ldots & S \\
64478 & Aa,Ab & 47605.7615 &     15.316 &      0.500 &     -0.165 &\ldots     & \ldots & \ldots & S \\
64478 & Aa,Ab & 47606.8573 &     15.291 &      0.500 &     -0.190 &\ldots     & \ldots & \ldots & S \\
64478 & Aa,Ab & 47607.9530 &     15.433 &      0.500 &     -0.048 &\ldots     & \ldots & \ldots & S \\
64478 & Aa,Ab & 47608.6835 &     15.494 &      0.500 &      0.012 &\ldots     & \ldots & \ldots & S \\
64478 & Aa,Ab & 47609.7792 &     16.075 &      0.500 &      0.593 &\ldots     & \ldots & \ldots & S \\
64478 & Aa,Ab & 57162.6889 & \ldots     & \ldots & \ldots &    28.004 &      0.250 &      0.039 & C \\
64478 & Aa,Ab & 57162.6889 & \ldots     & \ldots & \ldots &    27.628 &      0.250 &     -0.337 & C \\
64478 & Aa,Ab & 57166.7066 & \ldots     & \ldots & \ldots &    27.987 &      0.250 &      0.021 & C \\
64478 & Aa,Ab & 57166.7066 & \ldots     & \ldots & \ldots &    28.273 &      0.250 &      0.307 & C \\
64478 & Aa,Ab & 57169.2633 &     15.360 &  0.200 &  0.029 &    28.413 &      0.250 &      0.446 & C \\
64478 & Aa,Ab & 57175.4724 & \ldots     & \ldots & \ldots &    27.892 &      0.250 &     -0.077 & C \\
64478 & Aa,Ab & 57177.6638 & \ldots     & \ldots & \ldots &    27.944 &      0.250 &     -0.025 & C \\
64478 & Aa,Ab & 57177.6638 & \ldots     & \ldots & \ldots &    27.627 &      0.250 &     -0.342 & C \\
64478 & Ba,Bb & 57040.8218 &     48.951 &      0.500 &     -0.145 &     -44.695 &      2.000 &     -1.286 & C \\
64478 & Ba,Bb & 57078.7886 &     44.359 &      0.500 &     -0.445 &     -32.636 &      2.000 &      1.609 & C \\
64478 & Ba,Bb & 57093.7293 &     20.882 &     20.000 &      3.968 &  \ldots     &     \ldots &     \ldots & C \\
64478 & Ba,Bb & 57098.7453 &     40.107 &      0.500 &      1.422 &     -16.396 &      2.000 &      4.786 & C \\
64478 & Ba,Bb & 57120.6680 &     41.426 &      0.500 &     -0.206 &     -26.514 &      2.000 &      0.958 & C \\
64478 & Ba,Bb & 57121.7598 &      1.572 &     20.000 &      1.846 &      60.377 &      2.000 &     -1.617 & C \\
64478 & Ba,Bb & 57125.5607 &     49.006 &      0.500 &      0.232 &     -44.507 &      2.000 &     -1.786 & C \\
64478 & Ba,Bb & 57130.5543 &    -10.122 &     20.000 &     -0.385 &      83.650 &     20.000 &      1.453 & C \\
64478 & Ba,Bb & 57131.5839 &     24.598 &     20.000 &     12.628 &  \ldots     &     \ldots &     \ldots & C \\ 
64478 & Ba,Bb & 57134.6295 &     15.717 &     20.000 &    -10.357 &  \ldots     &     \ldots &     \ldots & C \\ 
64478 & Ba,Bb & 57137.7357 &     48.745 &      0.500 &      0.138 &     -43.823 &      2.000 &     -1.459 & C \\
64478 & Ba,Bb & 57141.6503 &     47.531 &      0.500 &     -0.449 &     -41.768 &      2.000 &     -0.742 & C \\
64478 & Ba,Bb & 57162.5514 &     39.797 &      0.500 &     -0.251 &     -21.691 &      2.000 &      2.401 & C \\
64478 & Ba,Bb & 57162.6074 &     42.843 &      0.500 &     -0.461 &     -30.154 &      2.000 &      0.888 & C \\
64478 & Ba,Bb & 57169.5361 &     -8.785 &      0.500 &     -0.472 &      80.482 &      2.000 &      1.326 & C \\
64478 & Ba,Bb & 57169.5378 &     -7.794 &      0.500 &      0.066 &      79.113 &      2.000 &      0.924 & C \\
64478 & Ba,Bb & 57169.5706 &     25.404 &     20.000 &     15.822 &  \ldots     &     \ldots &     \ldots & C \\ 
64478 & Ba,Bb & 57170.5193 &     -4.710 &      0.500 &      0.577 &      72.806 &      2.000 &      0.109 & C \\
64478 & Ba,Bb & 57175.6170 &    -10.108 &      0.500 &     -0.476 &      83.218 &      2.000 &      1.246 & C \\
64478 & Ba,Bb & 57177.5015 &     21.481 &     10.000 &      4.122 &   \ldots    &     \ldots &     \ldots & C \\
64478 & Ba,Bb & 57177.6205 &     19.332 &     10.000 &     -0.382 &   \ldots    &     \ldots &     \ldots & C 
\enddata 
\tablenotetext{a}{
C: CHIRON;
F: FECH;
S: \citet{Saar1990};
H: HARPS; 
L: \citet{survey}.
}
\end{deluxetable*}

\begin{deluxetable*}{r l l rrr rr l}    
\tabletypesize{\scriptsize}     
\tablecaption{Position measurements and residuals
\label{tab:speckle}          }
\tablewidth{0pt}                                   
\tablehead{                                                                     
\colhead{HIP} & 
\colhead{System} & 
\colhead{Date} & 
\colhead{$\theta$} & 
\colhead{$\rho$} & 
\colhead{$\sigma$} & 
\colhead{(O$-$C)$_\theta$ } & 
\colhead{(O$-$C)$_\rho$ } &
\colhead{Ref.\tablenotemark{a}} \\
 & & 
\colhead{(yr)} &
\colhead{(\degr)} &
\colhead{(\arcsec)} &
\colhead{(\arcsec)} &
\colhead{(\degr)} &
\colhead{(\arcsec)} &
}
\startdata
9642 & A,B &  1932.9100 &    177.9 &    1.020 &    0.500 &     34.9 &   -0.196 & Vis  \\
9642 & A,B &  1935.6600 &    159.6 &    1.170 &    0.050 &     13.1 &   -0.042 & Vis \\
9642 & A,B &  1936.9500 &    153.8 &    0.960 &    0.200 &      5.7 &   -0.251 & Vis \\
9642 & A,B &  1956.8600 &    165.9 &    0.960 &    0.200 &     -8.0 &   -0.233 & Vis  \\
9642 & A,B &  1991.2500 &    218.0 &    1.332 &    0.010 &      1.6 &    0.030 & HIP  \\
9642 & A,B &  2008.5407 &    233.7 &    1.425 &    0.002 &     -0.1 &   -0.004 & Spe  \\
9642 & A,B &  2008.7700 &    233.9 &    1.427 &    0.002 &     -0.1 &   -0.004 & SOAR  \\
9642 & A,B &  2011.0393 &    236.0 &    1.448 &    0.002 &      0.0 &   -0.002 & SOAR  \\
9642 & A,B &  2011.0393 &    236.1 &    1.448 &    0.002 &      0.0 &   -0.002 & SOAR \\
9642 & A,B &  2012.9205 &    237.6 &    1.463 &    0.002 &     -0.1 &   -0.002 & SOAR \\
9642 & A,B &  2012.9205 &    237.6 &    1.462 &    0.002 &     -0.1 &   -0.004 & SOAR  \\
9642 & A,B &  2012.9205 &    237.6 &    1.463 &    0.002 &     -0.1 &   -0.002 & SOAR  \\
9642 & A,B &  2015.0284 &    239.6 &    1.488 &    0.002 &      0.1 &    0.005 & SOAR \\
9642 & A,B &  2015.0284 &    239.6 &    1.487 &    0.002 &      0.1 &    0.003 & SOAR \\
9642 & A,B &  2015.1050 &    239.4 &    1.490 &    0.002 &     -0.1 &    0.006 & SOAR \\
9642 & A,B &  2015.7383 &    240.3 &    1.494 &    0.002 &      0.2 &    0.005 & SOAR \\
9642 & A,B &  2015.9131 &    240.3 &    1.490 &    0.002 &      0.1 &   -0.001 & SOAR \\
12780 & Aa,Ab &  1963.0500 &    142.7 &   0.1140 &   1.0500 &      3.5 &   0.0324 & Fin \\
12780 & Aa,Ab &  1964.0340 &    171.4 &   0.1400 &   0.0500 &      1.5 &   0.0097 & Fin \\
12780 & Aa,Ab &  1965.0460 &    185.4 &   0.1380 &   0.0500 &     -1.9 &  -0.0109 & Fin \\
12780 & Aa,Ab &  1966.0200 &    173.1 &   0.1400 &   1.0500 &    -30.1 &   0.0009 & Vis \\
12780 & Aa,Ab &  1966.0620 &    209.1 &   0.1280 &   0.0500 &      5.2 &  -0.0101 & Fin \\
12780 & Aa,Ab &  1967.1530 &    230.2 &   0.1290 &   0.0500 &     -0.9 &   0.0320 & Fin \\
12780 & Aa,Ab &  1978.9700 &    193.1 &   0.1700 &   0.0500 &     -2.7 &   0.0233 & Vis  \\
12780 & Aa,Ab &  1989.9330 &    146.0 &   0.0890 &   0.0020 &      1.5 &   0.0007 & Spe  \\
12780 & Aa,Ab &  1990.9130 &    172.5 &   0.1340 &   0.0020 &      0.4 &   0.0004 & Spe \\
12780 & Aa,Ab &  1990.9240 &    172.8 &   0.1340 &   0.0020 &      0.5 &   0.0000 & Spe \\
12780 & Aa,Ab &  1991.7210 &    184.2 &   0.1460 &   0.0020 &     -1.7 &  -0.0024 & Spe \\
12780 & Aa,Ab &  1991.7240 &    186.1 &   0.1490 &   0.0020 &      0.2 &   0.0006 & Spe \\
12780 & Aa,Ab &  2008.6060 &    339.1 &   0.0510 &   0.0020 &     -1.3 &  -0.0004 & SOAR \\
12780 & Aa,Ab &  2008.6060 &    339.2 &   0.0510 &   0.0020 &     -1.2 &  -0.0004 & SOAR   \\
12780 & Aa,Ab &  2009.6710 &    122.2 &   0.0650 &   0.0020 &     -2.3 &  -0.0019 & SOAR   \\
12780 & Aa,Ab &  2009.7560 &    130.1 &   0.0730 &   0.0020 &     -0.5 &   0.0008 & SOAR   \\
12780 & Aa,Ab &  2010.9660 &    171.8 &   0.1340 &   0.0020 &      0.2 &   0.0012 & SOAR   \\
12780 & Aa,Ab &  2013.7480 &    223.0 &   0.1100 &   0.0020 &      0.2 &   0.0008 & SOAR   \\
12780 & Aa,Ab &  2014.7690 &    279.1 &   0.0620 &   0.0020 &      1.8 &   0.0017 & SOAR   \\
12780 & Aa,Ab &  2014.7690 &    280.9 &   0.0600 &   0.0020 &      3.6 &  -0.0003 & SOAR   \\
12780 & Aa,Ab &  2015.0280 &    304.4 &   0.0520 &   0.0020 &     -1.3 &  -0.0014 & SOAR   \\
64478 & Aa,Ab &  1991.2500 &     96.0 &   0.1870 &   0.0100 &     -5.0 &  -0.0015 & HIP \\
64478 & Aa,Ab &  2001.0820 &    277.1 &   0.2420 &   0.0050 &     -3.3 &  -0.0021 & Spe \\
64478 & Aa,Ab &  2011.0400 &     98.8 &   0.2760 &   0.0020 &     -0.1 &  -0.0024 & SOAR \\
64478 & Aa,Ab &  2011.0400 &     98.4 &   0.2790 &   0.0020 &     -0.5 &   0.0006 & SOAR   \\
64478 & Aa,Ab &  2012.1020 &     99.2 &   0.3160 &   0.0020 &      0.1 &   0.0018 & SOAR   \\
64478 & Aa,Ab &  2014.3010 &     99.6 &   0.3560 &   0.0020 &      0.1 &   0.0014 & SOAR   \\
64478 & Aa,Ab &  2015.2500 &     99.9 &   0.3550 &   0.0020 &      0.3 &  -0.0009 & SOAR   \\
64478 & Aa,Ab &  2015.2500 &     99.8 &   0.3560 &   0.0020 &      0.2 &   0.0001 & SOAR   
\enddata 
\tablenotetext{a}{
Fin: ocular interferometry by W. S. Finsen;
HIP: Hipparcos;
SOAR: speckle interferometry at SOAR;
Spe: speckle interferometry at other telescopes;
Vis: visual micrometer measures.
}
\end{deluxetable*}

\end{document}